%% file: rt.tex
\documentclass[twoside,leqno,twocolumn]{article}  
\usepackage{amsmath}
\usepackage{amssymb}
\usepackage{amsthm}
\usepackage[cm]{fullpage}
\usepackage{graphicx}
\usepackage{pict2e}
\usepackage{subfig}
\usepackage[compact]{titlesec}
\titlespacing{\section}{0pt}{2ex}{1ex}
\titlespacing{\subsection}{0pt}{1ex}{0ex}
\usepackage{url}

\DeclareMathOperator*{\E}{\mathbf{E}}
\newcommand{\dist}[3]{\mathop{\textup{dist}}_{#1}(#2, #3)}
\newcommand{\NW}{N_W}
\newcommand{\NE}{N_E}
\newcommand{\SW}{S_W}
\newcommand{\SE}{S_E}

\newtheorem{theorem}{Theorem}
\newtheorem{lemma}[theorem]{Lemma}

\newtheorem{proposition}[theorem]{Proposition}

\title{Random road networks: the quadtree model}
\author{David Eisenstat\thanks{Department of Computer Science, Brown University.
Supported in part by NSF grant CCF-0964037.
\texttt{eisenstatdavid@gmail.com}}}
\date{}
\begin{document}
\maketitle
\begin{abstract}
\small
\baselineskip=9pt
What does a typical road network look like?
Existing generative models tend to focus on one aspect to the exclusion of others.
We introduce the general-purpose \emph{quadtree model} and analyze its shortest paths and maximum flow.
\end{abstract}
\section{Introduction}
\label{section:introduction}
\input{introduction.tex}
\section{Formal model description and summary of results}
\label{section:model-results}
\input{model-results.tex}
\section{Continuous-time reformulation}
\label{section:continuous-time}
\input{continuous-time.tex}
\section{Navigation (Proof of Theorem~\ref{theorem:navigation})}
\label{section:navigation}
\input{navigation.tex}
\section{Maximum flow}
\label{section:flow}
\input{flow.tex}
\section{Maximum flow when $r < 1$ (Proof of Theorem~\ref{theorem:flow-less-than-one})}
\label{section:flow-less-than-one}
\input{flow-less-than-one.tex}
\section{Maximum flow when $r = 1$ (Proof of Theorem~\ref{theorem:flow-equal-to-one})}
\label{section:flow-equal-to-one}
\input{flow-equal-to-one.tex}
\section{Maximum flow when $r > 1$ (Proof of Theorem~\ref{theorem:flow-greater-than-one})}
\label{section:flow-greater-than-one}
\input{flow-greater-than-one.tex}
\section{Acknowledgment}
The author would like to thank his advisor Claire Mathieu for much guidance and encouragement and the anonymous reviewers for their suggestions.
\bibliographystyle{plain}
\bibliography{rt}
\pagebreak
\section*{Appendix}
\label{section:appendix}
\input{appendix.tex}
\end{document}

%% file: introduction.tex
Many optimization problems naturally feature a road network.
Computing driving directions can be viewed as a shortest-path problem, and at a higher level, a typical vehicle routing problem is to schedule a fleet of capacity-limited trucks making deliveries from a depot to several customers.
Algorithms that solve these kinds of problems on general graphs are often impractical, leading algorithm designers to consider the special structure of road networks.
How can we characterize this structure?

Viewing the US road network as a metric space, Bast, Funke, and Matijevic found a small set of \emph{transit nodes} that covers almost all of the shortest paths~\cite{Bast}.
The existence of such a set underlies their point-to-point shortest-paths algorithm and was later formalized by Abraham, Fiat, Goldberg, and Werneck as \emph{highway dimension} in order to explain the performance of several competing algorithms~\cite{Abraham}.
Kalapala, Sanwalani, Clauset, and Moore examined several national road networks and discovered evidence of scale invariance~\cite{Kalapala}.
Aldous has explored the axiomatic consequences of scale invariance in road networks~\cite{Aldous}.
Masucci, Smith, Crooks, and Batty studied the London road network, finding, among other things, that the length of a shortest path is usually competitive with the distance between its endpoints and that most pairs of points are connected by a path with few turns~\cite{Masucci}.

Topologically, road networks have few crossings involving roads at different elevations, leading many researchers, including Masucci et al., to view road networks as planar graphs.
Eppstein and Goodrich, however, dispute both this characterization and a narrower one that excludes highways, citing evidence that in the US road network, crossings between local roads number on the order of $\sqrt n$, where $n$ is the number of intersections~\cite{Eppstein}.
They introduce as an alternative the notion of \emph{multiscale-dispersed} graphs.
Algorithmically, one key property of planar graphs~\cite{Lipton} and multiscale-dispersed graphs~\cite{Eppstein} is that they have \emph{separators} of size $O(\sqrt n)$, whose removal leaves two subgraphs of approximately the same size disconnected from one another.
Frederickson was the first of many to design separator-exploiting graph algorithms~\cite{Frederickson}.
Other properties exclusive to planar graphs are exploited by algorithms such as the $O(n \log n)$ maximum $s t$--flow algorithm due to Borradaile and Klein~\cite{Borradaile}, which can be used for large-scale empirical experiments with the quadtree model.

Despite the role of road networks in practical optimization, there have been few attempts to synthesize these properties into a generative model.
Ideally, such a road network model would be realistic enough to yield new insights, yet mathematically tractable so as to be useful in validating new specialized algorithms.
Motivated by the success of generative models in the study of social networks~\cite{Mitzenmacher}, we introduce a new model of random road networks based on quadtrees.

\subsection{Existing models}

One natural proposal is the uniform random planar graph.
It is well studied and has efficient sampling algorithms~\cite{Fusy,Schaeffer}, but it has more high-degree nodes than real road networks~\cite{Masucci} and does not come with a geometric embedding.
Gerke, Schlatter, Steger, and Taraz analyze an incremental random planar graph process where uniform random edges that do not complete a forbidden subgraph are added to an initially empty graph~\cite{Gerke}.
It has the same drawbacks, and a variant where straight edges are added between nodes chosen uniformly at random in the unit square is considered and rejected by Masucci et al.\ as being too difficult to navigate~\cite{Masucci}.

Abraham et al.\ describe an incremental road network construction algorithm that takes as input a sequence of points in a metric space~\cite{Abraham}.
Sets of input points $C_0 \supseteq C_1 \supseteq \cdots$ are maintained in such a way that $C_i$ is a maximal set of points at distance greater than $2^i$ from one another.
For all $i$, roads are added between points $p, q \in C_i$ with $\dist{}{p}{q} \le 6 \cdot 2^i$.
For all $p \in C_{i - 1} - C_i$, a road is also added from $p$ to the closest point in $C_i$.
In essence, this algorithm creates local road networks at different scales, together with connections between the scales.
It is introduced as motivation for the authors' hypothesis that real road networks have small highway dimension and is not intended to capture the topology of a road network.

Kalapala et al.\ introduce a simple model of fractal road networks in order to study scale invariance~\cite{Kalapala}.
The quadtree model differs from this model in using a random tree whose distribution is controlled by a sprawl parameter, instead of a tree complete to a specified depth.
Visually, the road networks produced by the model of Kalapala et al.\ are too uniform to resemble real road networks.

Masucci et al.\ propose the Growing Random Planar Graph (GRPG) model, which is based on their statistical analysis of the London road network~\cite{Masucci}.
The GRPG model starts with a single vertex and constructs a planar-embedded tree incrementally.
At each step, a random parent vertex is chosen along with a random offset, and a child is placed offset from the parent as long as the new edge does not cross an existing edge.
Short edges are added to obtain the finished product.
This model seems difficult to analyze formally in light of its brute-force approach to ensuring that the graph is planar, and its creators make no attempt to do so.

\subsection{The quadtree model}

\begin{figure*}
\centering
\input{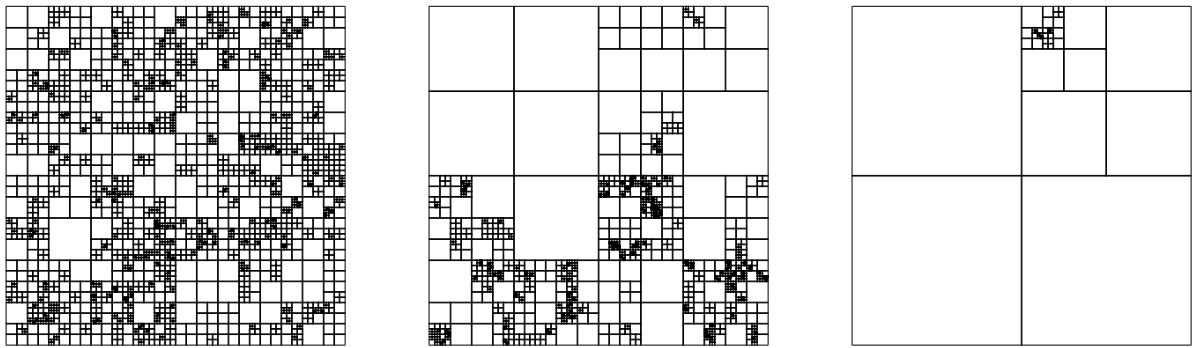}
\caption{the quadtree model.
Left: $r = 3/4$ (large-square preference).
Middle: $r = 1$ (neutral).
Right: $r = 5/4$ (small-square preference).}
\label{figure:example}
\end{figure*}

\begin{figure*}
\centering
\includegraphics[trim=0 0 72 72,width=128pt]{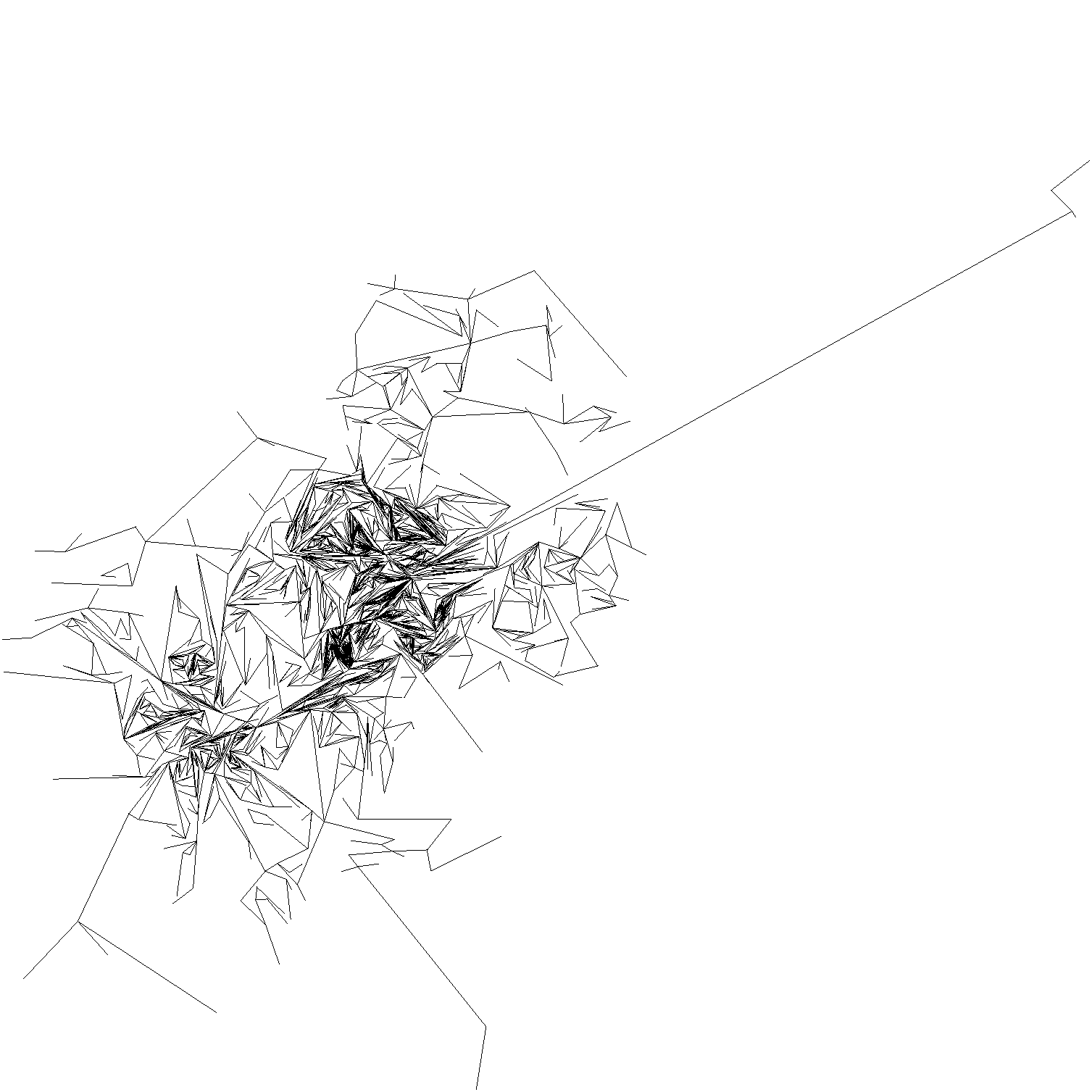}
\hspace{32pt}
\includegraphics[width=128pt]{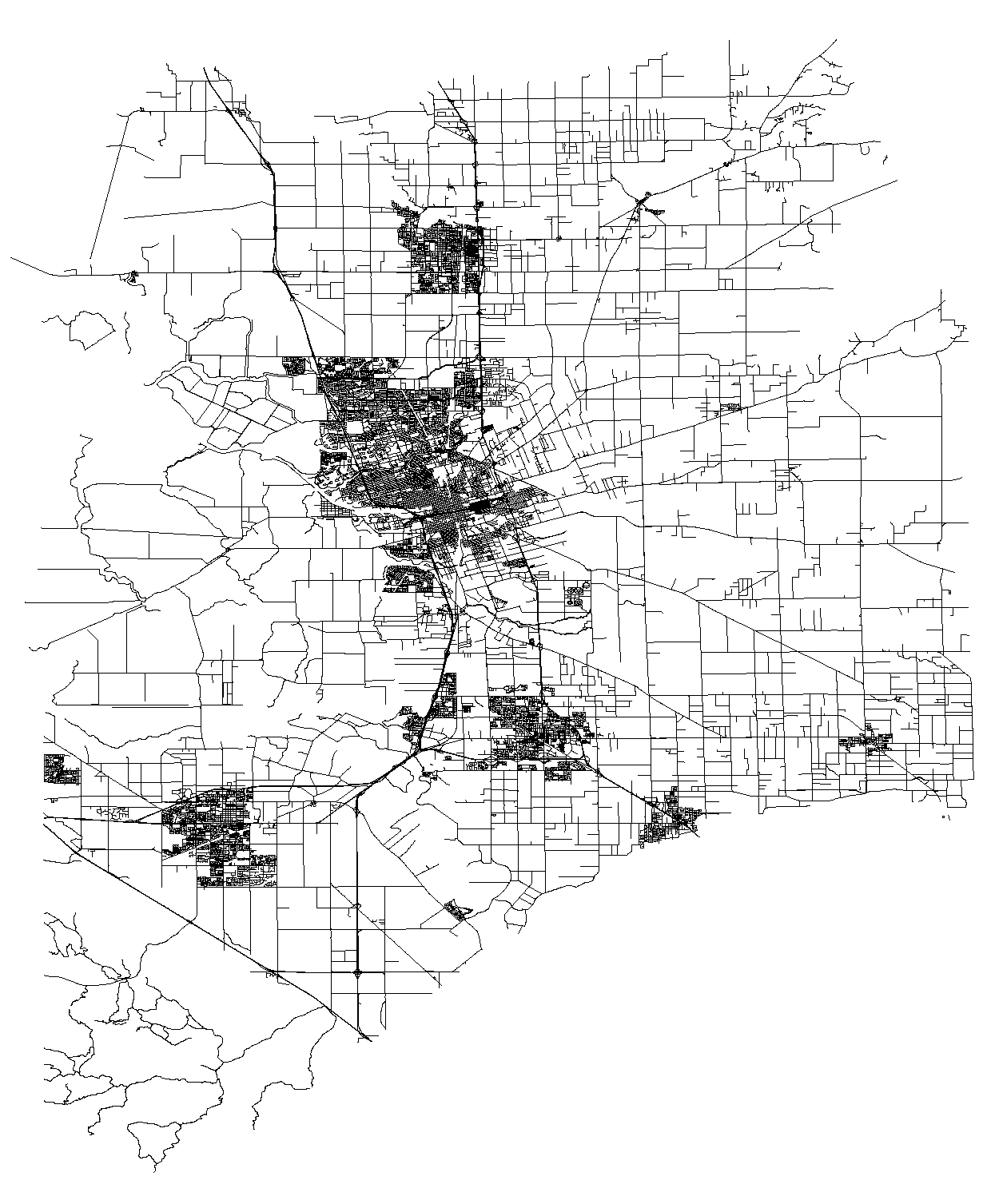}
\hspace{32pt}
\includegraphics[width=128pt]{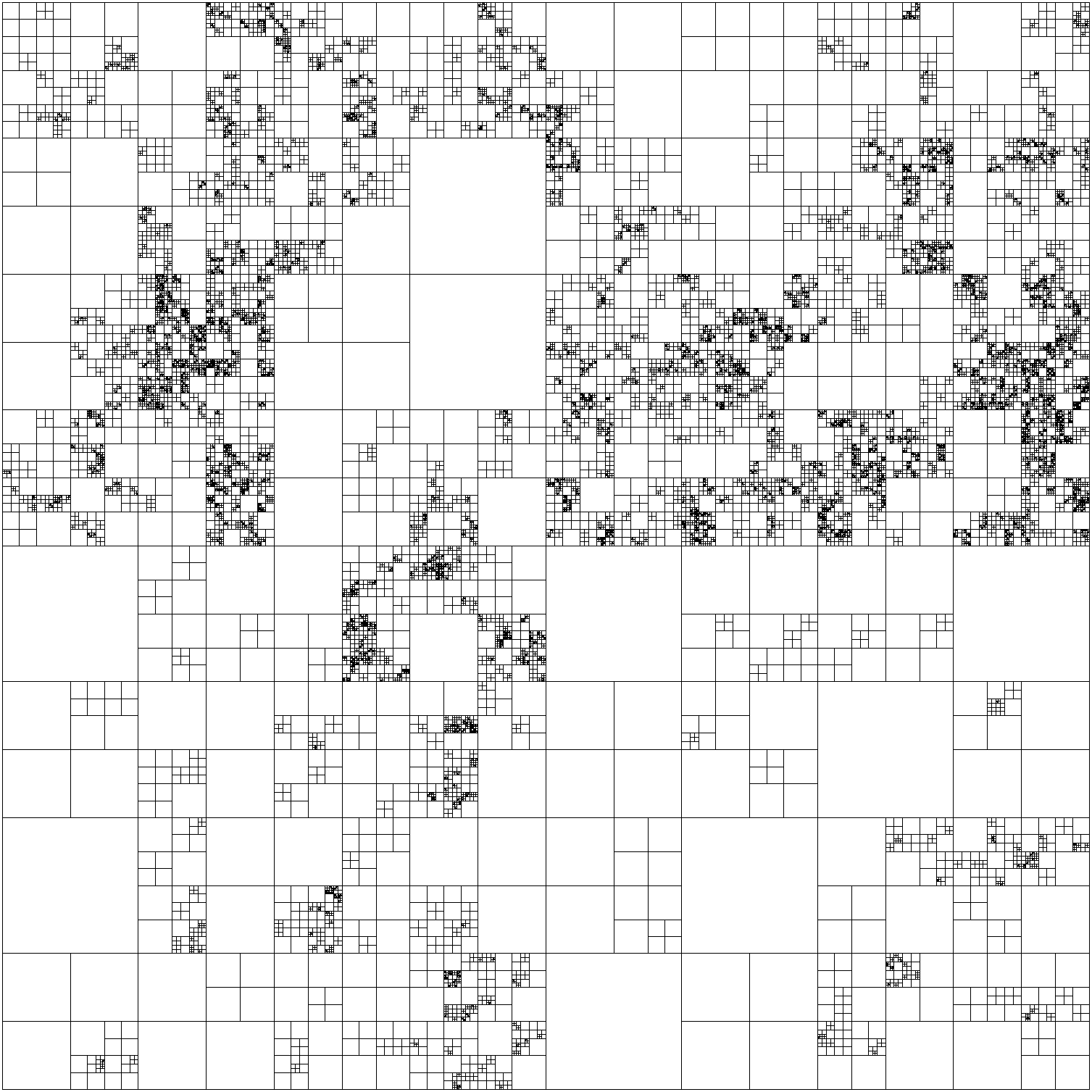}
\caption{Left: the Growing Random Planar Graph model~\cite{Masucci}.
Middle: the roads of San Joaquin County, California, USA~\cite{TIGER}.
Right: the quadtree model with $r = 15/16$.}
\label{figure:comparison}
\end{figure*}

The quadtree model has two parameters: $n$, the number of intersections, and $r$, which controls the amount of sprawl.
Imagine a square tract delimited by roads, to which $n$ people immigrate in succession.
Each new arrival chooses an existing square at random, builds two road segments bisecting it horizontally and vertically, and settles at the newly-created intersection.
This intersection is at the corner of four new delimited squares.
If people want to be far from other people ($r < 1$), then they prefer to settle larger squares.
Conversely, if they want to be close ($r > 1$), then they prefer smaller.
Figure~\ref{figure:example} shows the effects of different preferences.
Figure~\ref{figure:comparison} presents instances of the quadtree model and the GRPG model side by side with a real road network.


Road networks generated by the quadtree model have several desirable features.
They are embedded in the plane, self-similar, and for some choices of $r$, they display realistic variations in density.
Between any two points there exists a hierarchically organized path whose length is within a constant factor of their Euclidean distance and that with high probability when $r \le 1$ makes $O(\log n)$ turns (Theorem~\ref{theorem:navigation}; Section~\ref{section:navigation}).
This conforms with the observation of Masucci et al.\ that London is efficiently navigable in both senses~\cite{Masucci}.
Unlike the GRPG model, the quadtree model is simple and avoids an unrealistic abundance of acute-angled intersections.
Although the quadtree model does not specify speeds or capacities for roads, the technique of Kalapala et al.\ can be used to assign both based on the depth of each road in the tree~\cite{Kalapala}.
Different subdivision patterns can be used to avoid the monotony of square lots; we anticipate that our techniques will still apply with straightforward modifications.

\subsection{Maximum flow on road networks}

The capacity of road networks has received far less attention than shortest paths, yet for urban planners who must manage rush hour traffic and the possibility of natural disasters, it is a very important quantity.
Imagine that a large number of drivers wish to travel from the west edge of the road network to the east: how does sprawl affect the rate at which they succeed?

For simplicity, we assume that each road has unit capacity.
We anticipate that our techniques extend straightforwardly to more realistic settings where low-level roads have less capacity.
We measure the capacity of a road network by the maximum value of an east--west flow, which by our assumption equals the maximum number of edge-disjoint east--west paths.
By the max-flow min-cut theorem~\cite{Elias,Ford}, the capacity of a road network is also equal to the minimum total capacity of roads that when removed, disconnect the east from the west.

We obtain a partial characterization of maximum flow in the quadtree model by analyzing the capacity of a random vertical cut for upper bounds and by embedding a partial grid for lower bounds.
When $r < 1$, the road network asymptotically resembles a grid, with maximum flow on the order of $\sqrt{n}$ in expectation (Theorem~\ref{theorem:flow-less-than-one}; Section~\ref{section:flow-less-than-one}).
When $r > 1$, the squares nest very deeply and the maximum flow is constant in expectation (Theorem~\ref{theorem:flow-greater-than-one}; Section~\ref{section:flow-greater-than-one}).
These results are proved by analyzing a continuous-time variant of the quadtree process with exponential random variables.
When $r = 1$, we have been able to prove only that a lower bound is $\Omega(n^{0.086\cdots})$ and an upper bound is $O(n^{1/3})$, confirming the existence of a phase transition (Theorem~\ref{theorem:flow-equal-to-one}; Section~\ref{section:flow-equal-to-one}).
The proofs of both bounds involve a generating function.
Experimentally, the true exponent lies somewhere between the two, but its exact identity is unknown.
(Note that the quadtree in this case is uniform random.)

Previously, Eppstein, Goodrich, and Trott studied the expected number of roads intersected by a random line, both theoretically (for multiscale-dispersed graphs) and experimentally~\cite{Eppstein2}.
They proved for arbitrary multiscale-dispersed graphs that this number is $O(\sqrt n)$ but found that for real road networks it is significantly smaller.
The same number for the quadtree model with $r = 1$ is $\Theta(n^{1/3})$, which together with visual evidence, suggests that $r \triangleq 1 - \epsilon$ may be a realistic choice.

%% file: model-results.tex
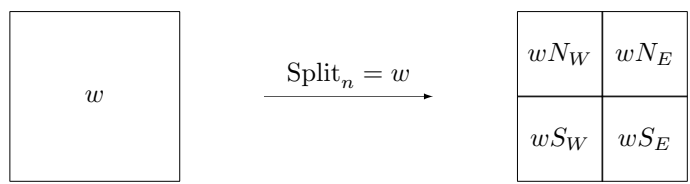
\begin{figure*}
\centering
\input{fig2}
\caption{Splitting square $w$.}
\label{figure:squares}
\end{figure*}

Let $\Sigma \triangleq \{\NW, \NE, \SW, \SE\}$ be an alphabet on four symbols.
We view strings as nodes in an infinite quadtree rooted at the empty string $\lambda$, where the children of each node $w \in \Sigma^*$ are $w a$ for all $a \in \Sigma$.
In turn, we view these nodes as squares.
Starting with the unit square $\lambda$, we cut each square horizontally and vertically to obtain its four children, as shown in Figure~\ref{figure:squares}.
Each square $w \in \Sigma^*$ so obtained is axis-aligned and has side length $2^{-|w|}$.

Let $r \in (0, \infty)$ be a parameter.
We specify a random growing road network as a sequence of sets of squares $\textup{Leaves}_n$.
Initially, there is only the unit square.
\begin{align*}
\textup{Leaves}_0 & \triangleq \{\lambda\}.
\end{align*}
For all integers $n \ge 0$, we choose a random square $\textup{Split}_n \in \textup{Leaves}_n$ to subdivide.
When $r < 1$, larger squares are more likely to be chosen.
When $r > 1$, smaller squares are more likely.
When $r = 1$, the choice is uniform.
The distribution of $\textup{Split}_n$ is given by
\begin{align*}
& \forall v \in \textup{Leaves}_n, \\
& \qquad \Pr(\textup{Split}_n = v \mid \textup{Leaves}_n) \triangleq \frac{r^{|v|}}{\sum_{w \in \textup{Leaves}_n} r^{|w|}}.
\end{align*}
To obtain the next set of squares, we replace $\textup{Split}_n$ by its children.
\begin{align*}
\textup{Leaves}_{n + 1} & \triangleq (\textup{Leaves}_n - \{\textup{Split}_n\}) \cup \{\textup{Split}_n a : a \in \Sigma\}.
\end{align*}
We call this model the \emph{quadtree model} for random road networks.
In the terminology of Bertoin~\cite[1.1]{Bertoin}, it is a conservative fragmentation chain with index of self-similarity $\alpha \triangleq \log_4(1/r)$.

For all integers $n \ge 0$, let $G_n$ be the graph of the square subdivision corresponding to $\textup{Leaves}_n$.
Let $\textup{Left}_n$ be the set of vertices with $x = 0$ and $\textup{Right}_n$ be the set of vertices with $x = 1$.
Define $\textup{MaxFlow}_n$ to be the maximum size of a set of pairwise edge-disjoint paths in $G_n$ from vertices in $\textup{Left}_n$ to vertices in $\textup{Right}_n$, that is, the maximum flow in $G_n$ from $\textup{Left}_n$ to $\textup{Right}_n$ assuming unit-capacity edges.

\begin{theorem}
\label{theorem:navigation}
Let $w, x \in \textup{Leaves}_n$ be squares, let $p$ be a point on the boundary of $w$, and let $q$ be a point on the boundary of $x$.
Define $v$ to be the least common ancestor of $w$ and $x$.
There exists a path in $G_n$ from $p$ to $q$ that changes direction at most $3 (d_w + d_x) + 6$ times and has length at most $2 \dist{1}{p}{q}$, where $d_w, d_x$ are the depths of $w, x$ in the subtree rooted at $v$.
\end{theorem}

\begin{theorem}
\label{theorem:flow-less-than-one}
If $r < 1$, then
\begin{align*}
\frac{n^{1/2}}{\exp O(\sqrt{\ln n})} & \le \E[\textup{MaxFlow}_n] \le O(n^{1/2}).
\end{align*}
\end{theorem}

The lower bound for $r < 1$ is very close to the upper bound; in this case, for all $\epsilon > 0$, we have $\E[\textup{MaxFlow}_n] = \Omega(n^{1/2 - \epsilon})$.

\begin{theorem}
\label{theorem:flow-equal-to-one}
If $r = 1$, then
\begin{align*}
\Omega(n^{0.086\cdots}) & \le \E[\textup{MaxFlow}_n] \le O(n^{1/3}).
\end{align*}
\end{theorem}

\begin{theorem}
\label{theorem:flow-greater-than-one}
If $r > 1$, then $\E[\textup{MaxFlow}_n] = \Theta(1)$.
\end{theorem}

%% file: fig2.tex
\begin{picture}(256,64)
\linethickness{0.25pt}
\put(0.00000000000000000,0.00000000000000000){\framebox(64.00000000000000000,64.00000000000000000){}}
\put(32.00000000000000000,32.00000000000000000){\makebox(0,0){$w$}}
\put(192.00000000000000000,32.00000000000000000){\framebox(32.00000000000000000,32.00000000000000000){}}
\put(224.00000000000000000,32.00000000000000000){\framebox(32.00000000000000000,32.00000000000000000){}}
\put(192.00000000000000000,0.00000000000000000){\framebox(32.00000000000000000,32.00000000000000000){}}
\put(224.00000000000000000,0.00000000000000000){\framebox(32.00000000000000000,32.00000000000000000){}}
\put(208.00000000000000000,48.00000000000000000){\makebox(0,0){$w\NW$}}
\put(240.00000000000000000,48.00000000000000000){\makebox(0,0){$w\NE$}}
\put(208.00000000000000000,16.00000000000000000){\makebox(0,0){$w\SW$}}
\put(240.00000000000000000,16.00000000000000000){\makebox(0,0){$w\SE$}}
\put(96.00000000000000000,32.00000000000000000){\vector(1,0){64}}
\put(128.00000000000000000,40.00000000000000000){\makebox(0,0){$\textup{Split}_n = w$}}
\end{picture}

%% file: continuous-time.tex
When $r \ne 1$, it is convenient to work with a continuous-time variant of the quadtree process.
In switching to continuous time, we lose fine control over the number of intersections but gain independence between disjoint subtrees.
The former is not an issue for us.

For all squares $w \in \Sigma^*$, let $\textup{Lifespan}(w)$ be an independent exponential random variable with rate $r^{|w|}$.
Define
\begin{align*}
\textup{Birth}(w) & \triangleq \sum_{v\textup{ is a proper ancestor of }w} \textup{Lifespan}(v) \\
\textup{Death}(w) & \triangleq \sum_{v\textup{ is an ancestor of }w} \textup{Lifespan}(v) \\
& = \textup{Birth}(w) + \textup{Lifespan}(w).
\end{align*}
For all times $t \in [0, \infty)$, let
\begin{align*}
\textup{Alive}(t) & \triangleq \{w : \textup{Birth}(w) \le t < \textup{Death}(w)\}.
\end{align*}
For all integers $n \ge 0$, let
\begin{align*}
T(n) & \triangleq \inf \{t : \#\textup{Alive}(t) = 3 n + 1\}.
\end{align*}
We assume, as is the case with probability $1$, that $T$ is everywhere defined.
The following proposition can be proved inductively by coupling the discrete process with the continuous process.

\begin{proposition}
For all integers $n \ge 0$, the random sets $\textup{Leaves}_n$ and $\textup{Alive}\bigl(T(n)\bigr)$ are identically distributed.
\end{proposition}

In the proofs that follow, we assume that $\textup{Leaves}_n = \textup{Alive}\bigl(T(n)\bigr)$.

%% file: navigation.tex
The key to short paths with few turns is using the hierarchical organization of the road network.
In essence, the proofs in this section work by replacing squares on a path in the quadtree with pieces of their boundaries (and possibly their siblings').
Note that when $r \le 1$, with high probability the quadtree has height $O(\log n)$.

\begin{figure*}
\centering
\input{fig3}
\caption{Left: Lemma~\ref{lemma:navigation} with $w \triangleq v \SW \NW$ and $p$ as shown.
Right: Theorem~\ref{theorem:navigation} with $w \triangleq v \SW \NW$ and $x \triangleq v \NE \SW$ and $p, q$ as shown.}
\label{figure:navigation-lemma}
\end{figure*}
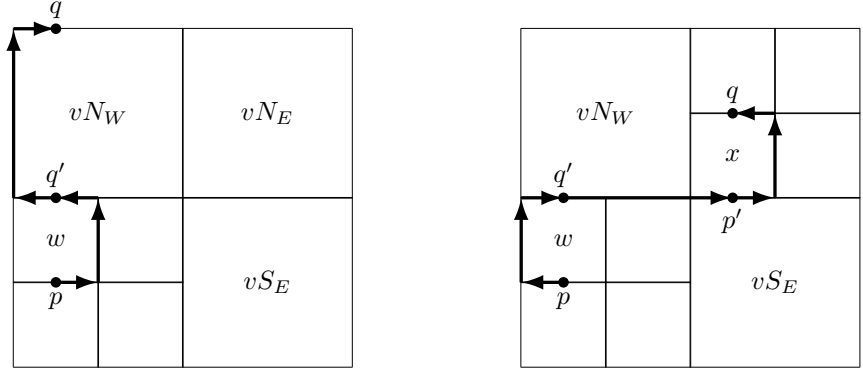

\begin{lemma}
Let $w \in \textup{Leaves}_n$ be a square.
Let $p$ be a point on the boundary of $w$ and let $v$ be an ancestor of $w$.
Define $q$ to be the unique point on the boundary of $v$ that is due north (respectively, west, east, south) of $p$.
There exists a path in $G_n$ from $p$ to $q$ that changes direction at most $3 d + 2$ times and has length at most $2 \dist{1}{p}{q}$, where $d$ is the depth of $w$ in the subtree rooted at $v$.
\label{lemma:navigation}
\end{lemma}
\begin{proof}
(See Figure~\ref{figure:navigation-lemma}.)
We induct on $d$.
If $d = 0$, then $w = v$, and $p$ is on the boundary of $v$.
Take the shortest path on the boundary of $v$.

If $d > 0$, then let $a \in \Sigma$ be the unique symbol such that $v' \triangleq v a$ is an ancestor of $w$.
Define $q'$ to be the unique point on the boundary of $v'$ that is due north of $p$.
By the inductive hypothesis, there exists a path in $G_n$ from $p$ to $q'$ that changes direction at most $3 d - 1$ times and has length at most $2 \dist{1}{p}{q'}$.
If $a \in \{\NW, \NE\}$, then $q' = q$, and no more travel is required.
Otherwise, from $q'$, we change directions at most $3$ more times in taking the shortest path to $q$ on the boundary of $v \NW$ (if $a = \SW$) or $v \NE$ (if $a = \SE$).
The total path length is at most $2 \dist{1}{p}{q'} + 2 \dist{1}{q'}{q} = 2 \dist{1}{p}{q}$, since $q'$ is on the line segment from $p$ to $q$.
\end{proof}

\begin{proof}[Proof of Theorem~\ref{theorem:navigation}]
Recall that $w, x \in \textup{Leaves}_n$ are squares whose least common ancestor is $v$.
The depths of $w, x$ in the subtree rooted at $v$ are $d_w, d_x$.
The point of departure $p$ lies on the boundary of $w$, and the point of arrival $q$ lies on the boundary of $x$.
Our goal is to find a path from $p$ to $q$ that changes direction at most $3 (d_w + d_x) + 6$ times and has length at most $2 \dist{1}{p}{q}$.

If $v = w = x$, then take the shortest path on the boundary of $v$.
Otherwise, $w, x \ne v$, since no square in $\textup{Leaves}_n$ has a proper ancestor in $\textup{Leaves}_n$.
Assume by rotational symmetry that $p$ is on or below the horizontal bisector of $v$ and that $q$ is on or above the bisector.
Let $q'$ be the unique point on the bisector due north of $p$ and let $p'$ be the unique point on the bisector due south of $q$.
Apply Lemma~\ref{lemma:navigation} to obtain paths from $p$ to $q'$ and from $p'$ to $q$.
Join these paths with a segment from $q'$ to $p'$ to obtain a path that changes directions at most $(3 d_w + 2) + 2 + (3 d_x + 2)$ times and has length at most $2 \dist{1}{p}{q'} + \dist{1}{q'}{p'} + 2 \dist{1}{p'}{q} \le 2 \dist{1}{p}{q}$.
\end{proof}

%% file: fig3.tex
\begin{picture}(320,128)
\linethickness{0.25pt}
\put(0.00000000000000000,64.00000000000000000){\framebox(64.00000000000000000,64.00000000000000000){}}
\put(64.00000000000000000,64.00000000000000000){\framebox(64.00000000000000000,64.00000000000000000){}}
\put(0.00000000000000000,32.00000000000000000){\framebox(32.00000000000000000,32.00000000000000000){}}
\put(32.00000000000000000,32.00000000000000000){\framebox(32.00000000000000000,32.00000000000000000){}}
\put(0.00000000000000000,0.00000000000000000){\framebox(32.00000000000000000,32.00000000000000000){}}
\put(32.00000000000000000,0.00000000000000000){\framebox(32.00000000000000000,32.00000000000000000){}}
\put(64.00000000000000000,0.00000000000000000){\framebox(64.00000000000000000,64.00000000000000000){}}
\put(32.00000000000000000,96.00000000000000000){\makebox(0,0){$v\NW$}}
\put(96.00000000000000000,96.00000000000000000){\makebox(0,0){$v\NE$}}
\put(16.00000000000000000,48.00000000000000000){\makebox(0,0){$w$}}
\put(48.00000000000000000,48.00000000000000000){\makebox(0,0){$$}}
\put(16.00000000000000000,16.00000000000000000){\makebox(0,0){$$}}
\put(48.00000000000000000,16.00000000000000000){\makebox(0,0){$$}}
\put(96.00000000000000000,32.00000000000000000){\makebox(0,0){$v\SE$}}
\linethickness{1.25pt}
\put(16.00000000000000000,32.00000000000000000){\vector(1,0){16}}
\put(32.00000000000000000,32.00000000000000000){\vector(0,1){32}}
\put(32.00000000000000000,64.00000000000000000){\vector(-1,0){16}}
\put(16.00000000000000000,64.00000000000000000){\vector(-1,0){16}}
\put(0.00000000000000000,64.00000000000000000){\vector(0,1){64}}
\put(0.00000000000000000,128.00000000000000000){\vector(1,0){16}}
\put(16.00000000000000000,32.00000000000000000){\circle*{4}}
\put(16.00000000000000000,64.00000000000000000){\circle*{4}}
\put(16.00000000000000000,128.00000000000000000){\circle*{4}}
\put(16.00000000000000000,28.00000000000000000){\makebox(0,0)[t]{$p$}}
\put(16.00000000000000000,68.00000000000000000){\makebox(0,0)[b]{$q'$}}
\put(16.00000000000000000,132.00000000000000000){\makebox(0,0)[b]{$q$}}
\linethickness{0.25pt}
\put(192.00000000000000000,64.00000000000000000){\framebox(64.00000000000000000,64.00000000000000000){}}
\put(256.00000000000000000,96.00000000000000000){\framebox(32.00000000000000000,32.00000000000000000){}}
\put(288.00000000000000000,96.00000000000000000){\framebox(32.00000000000000000,32.00000000000000000){}}
\put(256.00000000000000000,64.00000000000000000){\framebox(32.00000000000000000,32.00000000000000000){}}
\put(288.00000000000000000,64.00000000000000000){\framebox(32.00000000000000000,32.00000000000000000){}}
\put(192.00000000000000000,32.00000000000000000){\framebox(32.00000000000000000,32.00000000000000000){}}
\put(224.00000000000000000,32.00000000000000000){\framebox(32.00000000000000000,32.00000000000000000){}}
\put(192.00000000000000000,0.00000000000000000){\framebox(32.00000000000000000,32.00000000000000000){}}
\put(224.00000000000000000,0.00000000000000000){\framebox(32.00000000000000000,32.00000000000000000){}}
\put(256.00000000000000000,0.00000000000000000){\framebox(64.00000000000000000,64.00000000000000000){}}
\put(224.00000000000000000,96.00000000000000000){\makebox(0,0){$v\NW$}}
\put(272.00000000000000000,112.00000000000000000){\makebox(0,0){$$}}
\put(304.00000000000000000,112.00000000000000000){\makebox(0,0){$$}}
\put(272.00000000000000000,80.00000000000000000){\makebox(0,0){$x$}}
\put(304.00000000000000000,80.00000000000000000){\makebox(0,0){$$}}
\put(208.00000000000000000,48.00000000000000000){\makebox(0,0){$w$}}
\put(240.00000000000000000,48.00000000000000000){\makebox(0,0){$$}}
\put(208.00000000000000000,16.00000000000000000){\makebox(0,0){$$}}
\put(240.00000000000000000,16.00000000000000000){\makebox(0,0){$$}}
\put(288.00000000000000000,32.00000000000000000){\makebox(0,0){$v\SE$}}
\linethickness{1.25pt}
\put(208.00000000000000000,32.00000000000000000){\vector(-1,0){16}}
\put(192.00000000000000000,32.00000000000000000){\vector(0,1){32}}
\put(192.00000000000000000,64.00000000000000000){\vector(1,0){16}}
\put(208.00000000000000000,64.00000000000000000){\vector(1,0){64}}
\put(272.00000000000000000,64.00000000000000000){\vector(1,0){16}}
\put(288.00000000000000000,64.00000000000000000){\vector(0,1){32}}
\put(288.00000000000000000,96.00000000000000000){\vector(-1,0){16}}
\put(208.00000000000000000,32.00000000000000000){\circle*{4}}
\put(208.00000000000000000,64.00000000000000000){\circle*{4}}
\put(272.00000000000000000,64.00000000000000000){\circle*{4}}
\put(272.00000000000000000,96.00000000000000000){\circle*{4}}
\put(208.00000000000000000,28.00000000000000000){\makebox(0,0)[t]{$p$}}
\put(208.00000000000000000,68.00000000000000000){\makebox(0,0)[b]{$q'$}}
\put(272.00000000000000000,60.00000000000000000){\makebox(0,0)[t]{$p'$}}
\put(272.00000000000000000,100.00000000000000000){\makebox(0,0)[b]{$q$}}
\end{picture}

%% file: flow.tex
\begin{figure*}
\centering
\begin{minipage}{0.6\textwidth}
\renewcommand{\arraystretch}{1.75}
\begin{tabular}{cccc}
\hline
Case & Lower bound & Upper bound & Theorem \\
\hline\noalign{\smallskip}
$r < 1$ & $\frac{n^{1/2}}{\exp O(\sqrt{\ln n})}$ & $O(n^{1/2})$ & \ref{theorem:flow-less-than-one} \\
$r = 1$ & $\Omega(n^{0.086\cdots})$ & $O(n^{1/3})$ & \ref{theorem:flow-equal-to-one} \\
$r > 1$ & \multicolumn{2}{c}{$\Theta(1)$} & \ref{theorem:flow-greater-than-one} \\
\noalign{\smallskip}\hline
\end{tabular}
\renewcommand{\arraystretch}{1}
\end{minipage}
\begin{minipage}{0.3\textwidth}
\input{fig4}
\end{minipage}
\caption{Left: the asymptotic behavior of $\E[\textup{MaxFlow}_n]$ in $n$ for fixed $r$.
Right: Lemma~\ref{lemma:flow-lower} with $d \triangleq 3$.
Square $\NW \SE$ blocks $2^{3 - 2} - 1 = 1$ units of flow, and square $\SE$ blocks $2^{3 - 1} - 1 = 3$.}
\label{figure:flow-lower}
\end{figure*}
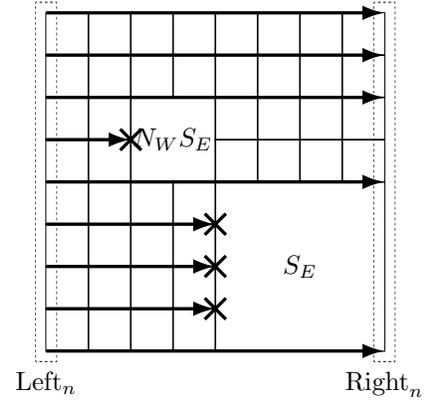

To bound $\textup{MaxFlow}_n$ below, we look to embed a regular grid in $G_n$.
When $r = 1$, it is advantageous to consider an incomplete grid, with \emph{holes}.
To bound $\textup{MaxFlow}_n$ above, we invoke the max-flow min-cut theorem and look for a vertical cut with low capacity.

For all integers $d \ge 0$, let
\begin{align*}
\textup{Holes}_n(d) & \triangleq \{w : w \in \textup{Leaves}_n,\: |w| < d\}
\end{align*}
be the set of \emph{holes} at depth $d$, that is, the set of squares that obstruct the embedding of a $2^d \times 2^d$ regular grid by having height larger than $2^{-d}$.

\begin{lemma}
\label{lemma:flow-lower}
For all integers $n, d \ge 0$,
\begin{align*}
\textup{MaxFlow}_n & \ge 2^d + 1 - \sum_{w \in \textup{Holes}_n(d)} (2^{d - |w|} - 1),
\end{align*}
\end{lemma}
\begin{proof}
(See Figure~\ref{figure:flow-lower}.)
We attempt to send $2^d + 1$ units of flow horizontally.
Each square $w \in \textup{Holes}_n(d)$ may block up to $2^{d - |w|} - 1$ units.
\end{proof}

\begin{lemma}
\label{lemma:flow-upper}
For all integers $n \ge 0$,
\begin{align*}
\textup{MaxFlow}_n & \le 1 + \sum_{w \in \textup{Leaves}_n} 2^{-|w|}.
\end{align*}
\end{lemma}
\begin{proof}
By the max-flow min-cut theorem~\cite{Elias,Ford}, every $\textup{Left}_n$--$\textup{Right}_n$ cut in the graph $G_n$ has capacity at least $\textup{MaxFlow}_n$.
Choose $U \in (0, 1) - \mathbf{Q}$ uniformly at random and form an $\textup{Left}_n$--$\textup{Right}_n$ cut by partitioning the vertices into those with $x < U$ and those with $x > U$ (all vertices have rational coordinates).
The capacity of this cut is equal to the number of squares intersected by the line $x = U$, plus one.
Each square $w \in \textup{Leaves}_n$ is intersected with probability equal to its width $2^{-|w|}$, and the conclusion follows by an averaging argument.
\end{proof}

Lemma~\ref{lemma:flow-upper} leads to an upper bound on $\textup{MaxFlow}_n$ that is tight when $G_n$ is a regular grid.

\begin{lemma}
\label{lemma:flow-max}
For all integers $n \ge 0$,
\begin{align*}
\textup{MaxFlow}_n & \le (3 n + 1)^{1/2} + 1.
\end{align*}
\end{lemma}
\begin{proof}
By Lemma~\ref{lemma:flow-upper}, it suffices to obtain an upper bound on the total square width plus one, which can be accomplished via the Cauchy--Schwarz inequality.
Let $\{w_1, \ldots, w_{3 n + 1}\} \triangleq \textup{Leaves}_n$ and consider the $(3 n + 1)$-dimensional vector $\mathbf{x}$ where $\mathbf{x}_k \triangleq 2^{-|w_k|}$.
Denoting by $\mathbf{1}$ the all-ones vector of the same dimension,
\begin{align*}
\sum_{w \in \textup{Leaves}_n} 2^{-|w|} & = \mathbf{1} \cdot \mathbf{x} \le \|\mathbf{1}\|_2\, \|\mathbf{x}\|_2 \\ & \qquad \textup{by Cauchy--Schwarz} \\
\|\mathbf{1}\|_2 & = (3 n + 1)^{1/2} \\
\|\mathbf{x}\|_2 & = \Bigl(\sum_{w \in \textup{Leaves}_n} 4^{-|w|}\Bigr)^{1/2} = 1 \\ & \qquad \textup{since }4^{-|w|}\textup{ is the area of }w.\tag*{\qedhere}
\end{align*}
\end{proof}

%% file: fig4.tex
\begin{picture}(128,152)
\linethickness{0.25pt}
\put(0.00000000000000000,128.00000000000000000){\framebox(16.00000000000000000,16.00000000000000000){}}
\put(16.00000000000000000,128.00000000000000000){\framebox(16.00000000000000000,16.00000000000000000){}}
\put(0.00000000000000000,112.00000000000000000){\framebox(16.00000000000000000,16.00000000000000000){}}
\put(16.00000000000000000,112.00000000000000000){\framebox(16.00000000000000000,16.00000000000000000){}}
\put(32.00000000000000000,128.00000000000000000){\framebox(16.00000000000000000,16.00000000000000000){}}
\put(48.00000000000000000,128.00000000000000000){\framebox(16.00000000000000000,16.00000000000000000){}}
\put(32.00000000000000000,112.00000000000000000){\framebox(16.00000000000000000,16.00000000000000000){}}
\put(48.00000000000000000,112.00000000000000000){\framebox(16.00000000000000000,16.00000000000000000){}}
\put(0.00000000000000000,96.00000000000000000){\framebox(16.00000000000000000,16.00000000000000000){}}
\put(16.00000000000000000,96.00000000000000000){\framebox(16.00000000000000000,16.00000000000000000){}}
\put(0.00000000000000000,80.00000000000000000){\framebox(16.00000000000000000,16.00000000000000000){}}
\put(16.00000000000000000,80.00000000000000000){\framebox(16.00000000000000000,16.00000000000000000){}}
\put(32.00000000000000000,80.00000000000000000){\framebox(32.00000000000000000,32.00000000000000000){}}
\put(64.00000000000000000,128.00000000000000000){\framebox(16.00000000000000000,16.00000000000000000){}}
\put(80.00000000000000000,128.00000000000000000){\framebox(16.00000000000000000,16.00000000000000000){}}
\put(64.00000000000000000,112.00000000000000000){\framebox(16.00000000000000000,16.00000000000000000){}}
\put(80.00000000000000000,112.00000000000000000){\framebox(16.00000000000000000,16.00000000000000000){}}
\put(96.00000000000000000,128.00000000000000000){\framebox(16.00000000000000000,16.00000000000000000){}}
\put(112.00000000000000000,128.00000000000000000){\framebox(16.00000000000000000,16.00000000000000000){}}
\put(96.00000000000000000,112.00000000000000000){\framebox(16.00000000000000000,16.00000000000000000){}}
\put(112.00000000000000000,112.00000000000000000){\framebox(16.00000000000000000,16.00000000000000000){}}
\put(64.00000000000000000,96.00000000000000000){\framebox(16.00000000000000000,16.00000000000000000){}}
\put(80.00000000000000000,96.00000000000000000){\framebox(16.00000000000000000,16.00000000000000000){}}
\put(64.00000000000000000,80.00000000000000000){\framebox(16.00000000000000000,16.00000000000000000){}}
\put(80.00000000000000000,80.00000000000000000){\framebox(16.00000000000000000,16.00000000000000000){}}
\put(96.00000000000000000,96.00000000000000000){\framebox(16.00000000000000000,16.00000000000000000){}}
\put(112.00000000000000000,96.00000000000000000){\framebox(16.00000000000000000,16.00000000000000000){}}
\put(96.00000000000000000,80.00000000000000000){\framebox(16.00000000000000000,16.00000000000000000){}}
\put(112.00000000000000000,80.00000000000000000){\framebox(16.00000000000000000,16.00000000000000000){}}
\put(0.00000000000000000,64.00000000000000000){\framebox(16.00000000000000000,16.00000000000000000){}}
\put(16.00000000000000000,64.00000000000000000){\framebox(16.00000000000000000,16.00000000000000000){}}
\put(0.00000000000000000,48.00000000000000000){\framebox(16.00000000000000000,16.00000000000000000){}}
\put(16.00000000000000000,48.00000000000000000){\framebox(16.00000000000000000,16.00000000000000000){}}
\put(32.00000000000000000,64.00000000000000000){\framebox(16.00000000000000000,16.00000000000000000){}}
\put(48.00000000000000000,64.00000000000000000){\framebox(16.00000000000000000,16.00000000000000000){}}
\put(32.00000000000000000,48.00000000000000000){\framebox(16.00000000000000000,16.00000000000000000){}}
\put(48.00000000000000000,48.00000000000000000){\framebox(16.00000000000000000,16.00000000000000000){}}
\put(0.00000000000000000,32.00000000000000000){\framebox(16.00000000000000000,16.00000000000000000){}}
\put(16.00000000000000000,32.00000000000000000){\framebox(16.00000000000000000,16.00000000000000000){}}
\put(0.00000000000000000,16.00000000000000000){\framebox(16.00000000000000000,16.00000000000000000){}}
\put(16.00000000000000000,16.00000000000000000){\framebox(16.00000000000000000,16.00000000000000000){}}
\put(32.00000000000000000,32.00000000000000000){\framebox(16.00000000000000000,16.00000000000000000){}}
\put(48.00000000000000000,32.00000000000000000){\framebox(16.00000000000000000,16.00000000000000000){}}
\put(32.00000000000000000,16.00000000000000000){\framebox(16.00000000000000000,16.00000000000000000){}}
\put(48.00000000000000000,16.00000000000000000){\framebox(16.00000000000000000,16.00000000000000000){}}
\put(64.00000000000000000,16.00000000000000000){\framebox(64.00000000000000000,64.00000000000000000){}}
\put(48.00000000000000000,96.00000000000000000){\makebox(0,0){$\NW\SE$}}
\put(96.00000000000000000,48.00000000000000000){\makebox(0,0){$\SE$}}
\put(-4.00000000000000000,12.00000000000000000){\dashbox(8.00000000000000000,136.00000000000000000){}}
\put(124.00000000000000000,12.00000000000000000){\dashbox(8.00000000000000000,136.00000000000000000){}}
\put(0.00000000000000000,8.00000000000000000){\makebox(0,0)[t]{$\textup{Left}_n$}}
\put(128.00000000000000000,8.00000000000000000){\makebox(0,0)[t]{$\textup{Right}_n$}}
\linethickness{1.25pt}
\put(60.00000000000000000,28.00000000000000000){\line(1,1){8}}
\put(60.00000000000000000,36.00000000000000000){\line(1,-1){8}}
\put(60.00000000000000000,44.00000000000000000){\line(1,1){8}}
\put(60.00000000000000000,52.00000000000000000){\line(1,-1){8}}
\put(60.00000000000000000,60.00000000000000000){\line(1,1){8}}
\put(60.00000000000000000,68.00000000000000000){\line(1,-1){8}}
\put(28.00000000000000000,92.00000000000000000){\line(1,1){8}}
\put(28.00000000000000000,100.00000000000000000){\line(1,-1){8}}
\put(0.00000000000000000,16.00000000000000000){\vector(1,0){128}}
\put(0.00000000000000000,32.00000000000000000){\vector(1,0){64}}
\put(0.00000000000000000,48.00000000000000000){\vector(1,0){64}}
\put(0.00000000000000000,64.00000000000000000){\vector(1,0){64}}
\put(0.00000000000000000,80.00000000000000000){\vector(1,0){128}}
\put(0.00000000000000000,96.00000000000000000){\vector(1,0){32}}
\put(0.00000000000000000,112.00000000000000000){\vector(1,0){128}}
\put(0.00000000000000000,128.00000000000000000){\vector(1,0){128}}
\put(0.00000000000000000,144.00000000000000000){\vector(1,0){128}}
\end{picture}

%% file: flow-less-than-one.tex
Given that large squares are much more likely to be split than small squares, it is probable that all square sizes lie in a narrow interval, in which case we can embed a relatively fine regular grid in $G_n$.
In fact, for any time $t$, we can say with some confidence what the sizes of the squares in $\textup{Alive}(t)$ will be.
Lemma~\ref{lemma:birth-lower} bounds the probability that a small square is born early.
\begin{lemma}
\label{lemma:birth-lower}
Assume that $r < 1$.
For all squares $w \in \Sigma^*$ and all integers $b \in [0, |w|]$,
\begin{align*}
\Pr\bigl(\textup{Birth}(w) < r^{-b}\bigr) & \le r^{\binom{|w| - b}{2}}.
\end{align*}
\end{lemma}
\begin{proof}
If $\textup{Birth}(w) < r^{-b}$, then for all proper ancestors $v$ of $w$, it holds that $\textup{Lifespan}(v) < r^{-b}$.
As $\textup{Lifespan}(v)$ is exponential with rate $r^{|v|}$, we have $\Pr\bigl( \textup{Lifespan}(v) < r^{-b} \bigr) = 1 - \exp(-r^{|v| - b}) \le \min\{r^{|v| - b}, 1\}$.
These events are independent, so we multiply their probabilities and obtain the stated bound.
\end{proof}

Lemma~\ref{lemma:death-upper} bounds the probability that a large square dies late.
\begin{lemma}
\label{lemma:death-upper}
Assume that $r < 1$.
For all squares $w \in \Sigma^*$ and all $p \in (0, 1]$,
\begin{align*}
\Pr \bigl( \textup{Death}(w) > \E[\textup{Death}(w)] \ln(1/p) \bigr) & \le (|w| + 1) p.
\end{align*}
\end{lemma}
\begin{proof}
If $\textup{Death}(w) > \E[\textup{Death}(w)] \ln(1/p)$, then there exists an ancestor $v$ of $w$ such that $\textup{Lifespan}(v) > \E[\textup{Lifespan}(v)] \ln(1/p)$.
For all $v$, the probability of the latter is $p$, and $w$ has $|w| + 1$ ancestors.
\end{proof}

Lemma~\ref{lemma:death-before-birth} establishes that with probability at least $1/2$, the last square at depth $c - O(\sqrt{c})$ dies before the first square at depth $c$ is born.
\begin{lemma}
\label{lemma:death-before-birth}
For all $r \in (0, 1)$, there exist constants $c_0, \alpha$ such that for all integers $c \ge c_0$, with probability at least $1/2$,
\begin{align*}
\forall v \in \Sigma^a,\: w \in \Sigma^c,\qquad \textup{Death}(v) & \le \textup{Birth}(w).
\end{align*}
where $a \triangleq c - \lfloor \alpha \sqrt{c} \rfloor$.
\end{lemma}
\begin{proof}
We exhibit a relatively late time ($r^{-b}$) at which with probability at least $3/4$, no square at depth $c$ has been born.
We then choose $a$ so that with probability at least $3/4$, every square at depth $a$ has been born.
More details are in the appendix.
\end{proof}

We now relate these results to $\textup{MaxFlow}_n$.

\begin{proof}[Proof of Theorem~\ref{theorem:flow-less-than-one}]
Lemma~\ref{lemma:flow-max} gives the upper bound $\E[\textup{MaxFlow}_n] = O(n^{1/2})$.
By an averaging argument, there exists a square $w \in \textup{Leaves}_n$ with area at most $1/(3 n + 1)$.
It thus holds that $|w| \ge c \triangleq \lceil \log_4(3 n + 1) \rceil$.
Defining $d \triangleq c - \lfloor \alpha \sqrt{c} \rfloor + 1$, where the constant $\alpha$ comes from Lemma~\ref{lemma:death-before-birth}, it holds with probability at least $1/2$ that $\textup{Holes}_n(d) = \{v : v \in \textup{Leaves}_n,\: |v| < d\} = \varnothing$.
We apply Lemma~\ref{lemma:flow-lower} and obtain the lower bound
\begin{align*}
\E[\textup{MaxFlow}_n] & \ge \frac{2^d + 1}{2} = \frac{n^{1/2}}{\exp O(\sqrt{\ln n})}.\tag*{\qedhere}
\end{align*}
\end{proof}

%% file: flow-equal-to-one.tex
For all integers $n \ge 0$, define an ordinary generating function
\begin{align*}
F_n(t) & \triangleq \sum_{w \in \textup{Leaves}_n} t^{|w|}.
\end{align*}
When $r = 1$, there is a nice closed form for $\E[F_n(t)]$.
This is helpful because using $F_n$, we can approximate from below the lower bound of Lemma~\ref{lemma:flow-lower} and express exactly the upper bound of Lemma~\ref{lemma:flow-upper}.

\begin{lemma}
\label{lemma:equal-to-one-F-upper}
Assume that $r = 1$.
For all integers $n \ge 0$ and all $t$,
\begin{align*}
\E[F_n(t)] & = \prod_{k = 0}^{n - 1} \Bigl(1 + \frac{4 t - 1}{3 k + 1}\Bigr) \le \exp\Bigl((4 t - 1) \sum_{k = 0}^{n - 1} \frac{1}{3 k + 1}\Bigr).
\end{align*}
\end{lemma}
\begin{proof}
For all integers $k \ge 0$,
\begin{align*}
\E[F_{k + 1}(t) \mid \textup{Leaves}_k] & = \Bigl(1 + \frac{4 t - 1}{3 k + 1}\Bigr) F_k(t).
\end{align*}
Since the set $\textup{Leaves}_k$ determines $F_k(t)$, we can condition on $F_k(t)$ alone.
The conclusion follows by induction and the inequality $1 + u \le \exp u$.
\end{proof}

\begin{proof}[Proof of Theorem~\ref{theorem:flow-equal-to-one}]
The upper bound is obtained by writing the total square width in terms of $F_n$.
By Lemma~\ref{lemma:flow-upper},
\begin{align*}
\textup{MaxFlow}_n & \le 1 + \sum_{w \in \textup{Leaves}_n} 2^{-|w|} = F_n(1/2) + 1.
\end{align*}
By Lemma~\ref{lemma:equal-to-one-F-upper} and integrating,
\begin{align*}
\E[\textup{MaxFlow}_n] & \le 1 + \exp\sum_{k = 0}^{n - 1} \frac{1}{3 k + 1} \\
& \le \exp\Bigl(1 + \frac{\ln(3 n - 2)}{3}\Bigr) + 1 = O(n^{1/3}).
\end{align*}
The lower bound is more involved.
To some extent, we emulate the proof of Devroye of the upper bound on the height of a random binary tree~\cite{Devroye}.
Let $c, d, t$ be defined by
\begin{align*}
c \ln(3 c/2) - c + 1/3 & \triangleq 1/\ln(3 n + 1) \\ & \qquad \bigl(c = 0.124\cdots - O(1/\ln n)\bigr) \\
d & \triangleq \lfloor c \ln(3 n + 1) \rfloor \\
t & \triangleq 3 c/4.
\end{align*}
We have
\begin{align*}
& \textup{MaxFlow}_n \\
& \ge 2^d + 1 - \sum_{w \in \textup{Holes}_n(d)} (2^{d - |w|} - 1) \\ & \qquad \textup{by Lemma~\ref{lemma:flow-lower}} \\
& > 2^d - \sum_{w \in \textup{Holes}_n(d)} (1/t)^{d - |w|} \\ & \qquad \textup{since }1/t \ge 2\textup{ and }\forall w \in \textup{Holes}_n(d),\:|w| \le d - 1 \\
& \ge 2^d - t^{-d} \sum_{w \in \textup{Leaves}_n} t^{|w|} \\ & \qquad \textup{since }\textup{Holes}_n(d) \subseteq \textup{Leaves}_n\textup{ and }t > 0 \\
& = 2^d - t^{-d} F_n(t).
\end{align*}
Take expectations, use Lemma~\ref{lemma:equal-to-one-F-upper}, integrate, and substitute $c, d, t$ (see the appendix).
\end{proof}

%% file: flow-greater-than-one.tex
Any square born after infinitely many other squares never enters the tree.
When $r > 1$, there exists with probability $1$ a least time $U \in [0, \infty)$ for which there exist infinitely many squares $v \in \Sigma^*$ with $\textup{Birth}(v) \le U$.
This effect is known as the explosion of Markov chains in continuous time.
At any given depth, most of the squares are born after time $U$, so the total width of all squares that ever enter the tree is small.

How can we bound $U$ above?
For all squares $v \in \Sigma^*$, define
\begin{align*}
\textup{Countdown}(v) & \triangleq \sum_{j = 0}^\infty \textup{Lifespan}(v \NW^j)
\end{align*}
to be the amount of time elapsed between the birth of $v$ and the time at which all of the infinitely many northwest descendants of $v$ have been born.
Given that $r > 1$, the sum converges with probability $1$, and $U \le \textup{Birth}(v) + \textup{Countdown}(v)$.

\begin{proposition}
\label{proposition:countdown}
For all squares $w \in \Sigma^*$, if there exists a square $v \in \Sigma^*$ such that $\textup{Birth}(v) + \textup{Countdown}(v) < \textup{Birth}(w)$, then for all integers $n \ge 0$, we have $w \notin \textup{Leaves}_n$.
\end{proposition}

For Proposition~\ref{proposition:countdown} to be useful, $\textup{Countdown}(v)$ must be small some of the time.
The significance of the constant $\ln(15/14)$ is that for all $w \in \Sigma^d$,
\begin{align*}
\Pr\bigl(\textup{Lifespan}(w) \le r^{-d} \ln(15/14)\bigr) & = 1/15.
\end{align*}

\begin{lemma}
\label{lemma:multiple-countdown}
Assume that $r > 1$.
There exists an integer $k \ge 1$ such that for all integers $d \ge 0$ and all sets of squares $V \subseteq \Sigma^d$ with $|V| = 4 k$, it holds with probability at least $15/16$ that there exists a square $v \in V$ such that
\begin{align*}
\textup{Countdown}(v) & \le r^{-d} \ln(15/14).
\end{align*}
\end{lemma}
\begin{proof}
See the appendix.
\end{proof}

For all integers $d \ge 0$, define
\begin{align*}
\textup{Born}_d & \triangleq \Sigma^d \cap \bigcup_{n = 0}^\infty \textup{Leaves}_n
\end{align*}
to be the set of all squares at depth $d$ that ever enter the tree.
In expectation, each set $\textup{Born}_d$ is small.

\begin{lemma}
\label{lemma:born}
Assuming that $r > 1$,
\begin{align*}
\E[\#\textup{Born}_d] = O(d).
\end{align*}
\end{lemma}
\begin{proof}
Rather than analyze $\#\textup{Born}_d$ directly, we define sequences of sets $\textup{Candidates}_d \supseteq \textup{Born}_d$ and $\textup{First}_d \subseteq \textup{Candidates}_d$.
The set $\textup{First}_d$ is chosen so that we can apply Lemma~\ref{lemma:multiple-countdown} with $V \triangleq \textup{First}_d$; the set $\textup{Candidates}_{d + 1}$ contains those squares that can still enter the tree given the quickest countdown in $\textup{First}_d$.
We prove that for all $d$,
\begin{align*}
\E[\#\textup{Candidates}_{d + 1} \mid \mathcal{F}_d] & \le 16 k (d + 1) + \#\textup{Candidates}_d/2,
\end{align*}
where the $\sigma$-algebra $\mathcal{F}_d$ belongs to a filtration that tracks the set of observed lifespans, and $k$ is the constant from Lemma~\ref{lemma:multiple-countdown}.
It follows that $\E[\#\textup{Candidates}_d] \le 32 k d$.

The sequences of sets $\textup{Candidates}_d$ and $\textup{First}_d$ are defined inductively.
Let
\begin{align*}
\textup{Candidates}_0 & \triangleq \{\lambda\}
\end{align*}
consist of the root square and let $\mathcal{F}_0$ be the trivial $\sigma$-algebra (condition on nothing).
The set $\textup{First}_d$, which is deterministic given $\mathcal{F}_d$, consists of the $4 k$ squares $v \in \textup{Candidates}_d$ with the least values of $\textup{Birth}(v)$ for which $\textup{Countdown}(v)$ is independent of $\mathcal{F}_d$.
(If there are fewer than $4 k$ such squares, take all of them.)
Let
\begin{align*}
& \textup{Parents}_d \triangleq \\
& \quad \{w : w \in \textup{Candidates}_d, \\
& \qquad \forall v \in \textup{First}_d,\: \textup{Death}(w) \le \textup{Birth}(v) + \textup{Countdown}(v)\}
\end{align*}
be a set of squares whose children can still enter the tree.
Finally, define
\begin{align*}
\textup{Candidates}_{d + 1} & \triangleq \{w a : w \in \textup{Parents}_d,\: a \in \Sigma\}
\end{align*}
and let $\mathcal{F}_{d + 1}$ be the $\sigma$-algebra generated by $\mathcal{F}_d$, the random variables $\textup{Lifespan}(v \NW^j)$ for $v \in \textup{First}_d$ and $j \ge 0$, and the random variables $\textup{Lifespan}(w)$ for $w \in \Sigma^d$.

By construction and Proposition~\ref{proposition:countdown}, it holds that $\textup{Born}_d \subseteq \textup{Candidates}_d$.
Of the squares in $\textup{Candidates}_d$, at most $4 k d$ have countdowns that are not independent of $\mathcal{F}_d$, one for each square in $\textup{First}_0 \cup \cdots \cup \textup{First}_{d - 1}$.
At most $4 k$ belong to $\textup{First}_d$.
As for the remainder, with probability at least $15/16$ there exists by Lemma~\ref{lemma:multiple-countdown} a square $v \in \textup{First}_d$ such that $\textup{Countdown}(v) \le r^{-d} \ln(15/14)$.
For all remaining squares $w$,
\begin{align*}
& \Pr(w \in \textup{Parents}_d \mid \mathcal{F}_d) \\
& \le \Pr\bigl(\textup{Birth}(w) + \textup{Lifespan}(w) \\
& \qquad \qquad \le \textup{Birth}(v) + \textup{Countdown}(v) \mid \mathcal{F}_d\bigr) \\
& \le \Pr\bigl(\textup{Lifespan}(w) \le \textup{Countdown}(v) \mid \mathcal{F}_d\bigr) \\ & \qquad \textup{since }\textup{Birth}(v) \le \textup{Birth}(w) \\
& \le 1/15,
\end{align*}
for a contribution of $\#\textup{Candidates}_d/4$ to $\E[\#\textup{Candidates}_{d + 1} \mid \mathcal{F}_d]$.
The event where $v$ does not exist contributes the same amount.
\end{proof}

\begin{proof}[Proof of Theorem~\ref{theorem:flow-greater-than-one}]
The lower bound $\textup{MaxFlow}_n = \Omega(1)$ is trivial.
As for the upper bound,
\begin{align*}
\textup{MaxFlow}_n & = 1 + \sum_{w \in \textup{Leaves}_n} 2^{-|w|} \\ & \qquad \textup{by Lemma~\ref{lemma:flow-upper}} \\
& \le 1 + \sum_{d = 0}^\infty \sum_{w \in \textup{Born}_d} 2^{-|w|} \\ & \qquad \textup{since }\textup{Leaves}_n \subseteq \bigcup_{d = 0}^\infty \textup{Born}_d \\
\E[\textup{MaxFlow}_n] & = 1 + \sum_{d = 0}^\infty 2^{-d} O(d) = O(1) \\ & \qquad \textup{by Lemma~\ref{lemma:born}.}\tag*{\qedhere}
\end{align*}
\end{proof}

%% file: appendix.tex
\begin{proof}[Proof of Lemma~\ref{lemma:death-before-birth}, continued]
Let $b \in [0, c]$ be the greatest integer such that
\begin{align*}
4^c r^{\binom{c - b}{2}} & \le 1/4.
\end{align*}
Let $a \in [0, b]$ be the greatest integer such that
\begin{align*}
\frac{r^{-a}}{1 - r} \ln \frac{1}{p(a)} & \le r^{-b},
\end{align*}
where $p(a) \triangleq 1/[4^{a + 1} (a + 1)]$.
Some analysis reveals that $b = c - O(\sqrt{c})$ and $a = b - O(\ln c)$.

Since $|\Sigma^c| = 4^c$, by Lemma~\ref{lemma:birth-lower}, it holds with probability at least $3/4$ that
\begin{align*}
\forall w \in \Sigma^c,\qquad \textup{Birth}(w) & \ge r^{-b},
\end{align*}
that is, that no birth at depth $c$ has occurred.
Observe that for all $v \in \Sigma^a$,
\begin{align*}
\E[\textup{Death}(v)] & = \sum_{k = 0}^{a} r^{-k} < \frac{r^{-a}}{1 - r}.
\end{align*}
By Lemma~\ref{lemma:death-upper} with $p \triangleq p(a)$, it holds with probability at least $3/4$ that
\begin{align*}
\forall v \in \Sigma^a,\qquad \textup{Death}(v) & \le r^{-b},
\end{align*}
that is, that every birth at depth $a$ has occurred.
\end{proof}

\begin{proof}[Proof of Theorem~\ref{theorem:flow-equal-to-one}, continued]
\begin{align*}
& \E[\textup{MaxFlow}_n] \\
& > 2^d - t^{-d} \E[F_n(t)] \\
& \ge 2^d - t^{-d} \exp\Bigl((4 t - 1) \sum_{k = 0}^{n - 1} \frac{1}{3 k + 1}\Bigr) \\ & \qquad \textup{by Lemma~\ref{lemma:equal-to-one-F-upper}} \\
& \ge 2^d - t^{-d} \exp\Bigl((4 t - 1) \frac{\ln(3 n + 1)}{3}\Bigr) \\ & \qquad \textup{by integrating, since }4 t - 1 < 0 \\
& = 2^d \Bigl(1 - (3 c/2)^{-d} \exp\bigl((c - 1/3) \ln(3 n + 1)\bigr)\Bigr) \\
& \ge 2^d \Bigl(1 - \exp\bigl([-c \ln(3 c/2) + c - 1/3] \ln(3 n + 1)\bigr)\Bigr) \\ & \qquad \textup{since }3 c/2 < 1\textup{ and }d \le c \ln(3 n + 1) \\
& = 2^d \bigl(1 - \exp(-1)\bigr) \\
& = \Omega(n^{0.086\cdots}).\tag*{\qedhere}
\end{align*}
\end{proof}

\begin{proof}[Proof of Lemma~\ref{lemma:multiple-countdown}]
Recalling that $\lambda$ is the empty string, let
\begin{align*}
u & \triangleq \ln(15/14) \\
p & \triangleq \Pr\bigl(\textup{Countdown}(\lambda) \le u\bigr).
\end{align*}
For all squares $w \in \Sigma^d$, we have $\Pr\bigl(\textup{Countdown}(w) \le r^{-d} \ln(15/14)\bigr) = p$.
After showing that $p > 0$, we finish by setting $k \ge 1$ to the least integer such that $(1 - p)^{4 k} \le 1/16$.

Define $k \ge 0$ to be the least integer such that
\begin{align*}
\sum_{j = k}^\infty r^{-j} & \le u/4.
\end{align*}
By Markov's inequality,
\begin{align*}
\Pr\Bigl(\sum_{j = k}^\infty \textup{Lifespan}(\NW^j) \le u/2\Bigr) & \ge 1/2.
\end{align*}
For all integers $j \in [0,\: k - 1]$, it holds with positive probability that $\textup{Lifespan}(\NW^j) \le u/(2 k)$, and all of the terms are independent.
\end{proof}